\documentclass[aps,prd,longbibliography,floatfix,twocolumn,groupedaddress,amssymb,nofootinbib]{revtex4-1}

\usepackage[latin1]{inputenc}
\usepackage{amsmath}
\usepackage{epstopdf}
\usepackage{flushend}   
\usepackage{subfigure}
\usepackage[title]{appendix}
\usepackage[hidelinks]{hyperref}
\usepackage{float}
\usepackage{graphicx}
\usepackage{dcolumn}
\usepackage{bm}

\usepackage{xcolor}
\hypersetup{
  colorlinks = true, 
  urlcolor = blue!50!black, 
  linkcolor = blue!50!black, 
  citecolor = blue!50!black 
}

\begin{document}
\title{Novel approach to assess the impact of the Fano factor on the sensitivity of low-mass dark matter experiments}
\author{D.~Durnford}
\email{d.durnford@queensu.ca}
\affiliation{Department of Physics, Engineering Physics $\&$ Astronomy, Queen's University, Kingston, Ontario K7L 3N6, Canada}


\author{Q.~Arnaud}
\affiliation{Department of Physics, Engineering Physics $\&$ Astronomy, Queen's University, Kingston, Ontario K7L 3N6, Canada}
\author{G.~Gerbier}
\affiliation{Department of Physics, Engineering Physics $\&$ Astronomy, Queen's University, Kingston, Ontario K7L 3N6, Canada}

\begin{abstract}
As first suggested by U. Fano in the 1940s, the statistical fluctuation of the number of pairs produced in an ionizing interaction is known to be sub-Poissonian. The dispersion is reduced by the so-called ``Fano factor," which empirically encapsulates the correlations in the process of ionization. In modeling the energy response of an ionization measurement device, the effect of the Fano factor is commonly folded into the overall energy resolution. While such an approximate treatment is appropriate when a significant number of  ionization pairs are expected to be produced, the Fano factor needs to be accounted for directly at the level of pair creation when only a few are expected. To do so, one needs a discrete probability distribution of the number of pairs created $N$ with independent control of both the expectation $\mu$ and Fano factor $F$. Although no distribution $P(N|\mu,F)$ with this convenient form exists, we propose the use of the COM-Poisson distribution together with strategies for utilizing it to effectively fulfill this need. We then use this distribution to assess the impact that the Fano factor may have on the sensitivity of low-mass WIMP search experiments.
\end{abstract}
\maketitle
\section{Introduction}
Following an ionizing particle interaction of deposited energy $E$, the mean number of pairs created is given by
\begin{equation}
\label{Np eq}
\mu = \frac{E}{W(E)},
\end{equation}
\noindent where $W$ is generally both energy and particle-type dependent and represents the mean energy needed to create either one electron-ion pair in liquid and gaseous detectors or one electron-hole pair in semiconductor devices \cite{sauli}. The actual number of pairs created $N$ is subject to statistical fluctuations, which limits the achievable energy resolution of any ionization measuring device to monoenergetic radiation. As first anticipated by Fano, the variance of these fluctuations $\sigma_N^2$ is lower than expected for a Poisson process by a factor $F$, known hereafter as the ``Fano factor," which is defined as \cite{uno2}
\begin{equation}
\label{Fdef eq}
F = \frac{\sigma_N^2}{\mu}.
\end{equation}
By definition $F=1$ for a Poissonian process, whereas $F<1$ for ionization fluctuations. Experimentally measuring the Fano factor is challenging, as one needs to both strongly suppress and precisely quantify all sources of resolution degradation that do not arise from ionization fluctuations. In spite of this, measurements of $F$ have been carried out for a wide variety of materials including argon ($F \approx 0.23$), xenon ($F \approx 0.17$), silicon ($F \approx 0.16$), germanium ($F \approx 0.12$), and others \cite{scinprop,policarpo,owens,germanium}.

Although these measurements set a fundamental upper limit on the resolution possible with these detector media, they do not provide more information about the actual probability distribution of the number of pairs created than its dispersion. While the latter can be predicted with Monte Carlo simulations of the processes involved in energy loss at a microscopic level \cite{gross1} (see Supplemental Material \cite{supp}), this approach is too computationally expensive to be practical for most applications. This is true for any scenario in which one needs to simulate the measurement of a signal that is not monoenergetic. In this case, one might think to fold the effect of the Fano factor into the overall energy resolution. While such an approach is appropriate at high energies, it is not valid when $\mu$ is small, in which case a more accurate treatment is necessary. To account for the Fano factor at the level of pair creation, one would require a discrete probability distribution $P \left( N | \mu,F \right)$ of the number $N$ of pairs created for any value of $\mu$ and $F$. Although there is no distribution with this exact convenient form, we propose the COM-Poisson distribution as a viable solution. It is a discrete distribution that allows for independent control of the mean and variance with two parameters, $\lambda$ and $\nu$. While these variables do not correspond to $\mu$ and $F$, we have developed a methodology to effectively translate $P \left( N | \lambda,\nu \right)$ into $P \left( N | \mu,F \right)$. 
\indent We believe the COM-Poisson distribution may provide a much needed tool in the area of low-mass dark matter research. Since new, popular models favor particle masses on the order of a few $\mathrm{GeV/c^2}$ or less \cite{Essig2013,ZUREK201491}, a growing cohort of direct detection experiments are now confronted with the issue of modeling ionization statistics at the single pair regime. This includes gaseous dark matter search experiments like NEWS-G \cite{Arnaud2018}, liquid noble experiments like DarkSide \cite{darkside1}, and solid-state experiments such as SuperCDMS, Edelweiss, DAMIC, and Sensei \cite{single,edelweiss,damic,sensei}. While these detector technologies differ in many ways, the requirements for modeling ionization statistics are essentially the same for each, and are fulfilled by the COM-Poisson distribution.

What follows is a more detailed discussion about the problem of modeling ionization statistics (Sec.\ \ref{model sec}), the COM-Poisson distribution (Sec.\ \ref{COM sec}), and strategies for using it (Sec.\ \ref{using sec}). Finally, we use the COM-Poisson distribution to assess the potential impact of the Fano factor on the sensitivity of dark matter detection experiments in Sec.\ \ref{App sec}.
\section{Modeling Ionization Statistics}
\label{model sec}
A detector's energy response to monoenergetic radiation is defined as the convolution of the probability distribution of ionization with the detector's energy resolution function. At high energies when one expects a large number of pairs to be created, the detector response tends to a Gaussian due to the central limit theorem \cite{Cowan1998}. Therefore the Fano factor can approximately be accounted for by including it in the standard deviation of the overall energy response. However, this approach is not appropriate at the single-pair regime, which particle detectors can now probe as experiments push the low-energy frontier. This, together with the improved understanding/reduction of other resolution degrading factors means that the Fano factor must be accounted for directly at the level of pair creation. Doing so in a Monte Carlo simulation would ideally require a probability distribution $P \left( N | \mu,F \right)$ to model the probability of creating $N$ pairs. We considered four minimal requirements for a probability distribution to be appropriate for this task:
\begin{enumerate}
\item That the probability distribution is discrete.
\item That it allows for independent control of both the mean and the Fano factor.
\item That it is defined for continuous values of the mean.
\item That it is defined for values of the Fano factor within an appropriate range. Specifically, we required that the distribution be defined for any value of $F$ that is $\leq 1$ down to $\sim 0.1$.
\end{enumerate}
This is not an exhaustive list of requirements for a model; one could consider other distribution shape properties such as kurtosis and skewness for example. However, in this work we concern ourselves only with these minimum requirements, as fulfilling even these is challenging. There are several well-known discrete probability distributions to consider for this purpose, as well as several potential candidates. One distribution first considered is the binomial distribution, with probability distribution function (PDF):
\begin{equation}
\begin{gathered}
\label{binom pdf eq}
P(X = k| n,p) = \binom nk p^k (1-p)^{n-k} \\\mathrm{for} \;\; n \in \mathbb{N}_0, \;\; k \leq n, \;\; p \in [0,1].
\end{gathered}
\end{equation}
It is a discrete distribution (it satisfies the first requirements), with a mean of $\mu = np$ (which satisfies the third requirement). The variance is given by $np(1-p)$, and so it does allow the mean and the Fano factor to be varied. However, we can express the Fano factor for the binomial distribution as
\begin{equation}
F = 1 - p = 1 - \frac{\mu}{n}.
\end{equation}
\label{binom eq}
\indent Thus, while the binomial distribution can be used with some values of $F$, it is not an appropriate model. Because $n$ is an integer, the distribution is not defined for continuous values of $F$ as we require, and the allowed values of $F$ vary with $\mu$ \cite{supp}. There are also many niche distributions designed specifically to satisfy the need for underdispersed or overdispersed models. Examples of these include the negative-binomial distribution, a good tool for overdispersion only \cite{count}. Another is the generalized Poisson distribution, which can also model underdispersion, but there is a lower limit on the Fano factor achievable with this distribution, so it is also inappropriate \cite{count,consul}. A class of distribution that can satisfy our requirements is the family of weighted Poisson distributions. This includes anything that can be written in the form \cite{weight}
\begin{equation}
\label{weight eq}
P\left(X = x | \lambda, w \right) = \frac{e^{-\lambda} \lambda^x w_x}{W x!} \quad \mathrm{for} \;\; x \in \mathbb{N}_0, \; \lambda > 0,
\end{equation}
\noindent where $w_x$ is a weight function (usually with two parameters itself) and $W$ is a normalizing constant. As a solution to the problem of modeling ionization statistics we propose the COM-Poisson distribution. It is a member of the family of weighted Poisson distributions, with weight function $w_x = \left(x! \right)^{1 - \nu}$ \cite{count}. While the distribution parameters $\lambda$ and $\nu$ do not correspond to the mean and the Fano factor, having only two parameters makes it easier to translate the COM-Poisson distribution $P \left(N | \lambda,\nu \right)$ into $P \left(N | \mu, F \right)$. Thus it is a more ``user-friendly" case of a weighted Poisson distribution. Additionally, our choice of the COM-Poisson distribution benefits from studies of its properties by others \cite{usefulCOM,mainCOM,compute}, many of which we make use of (see Sec.\ \ref{COM sec}). It meets all of our requirements: it is discrete, allows for independent control of the mean $\mu$ and $F$, and is defined for continuous values of $\mu$ and $F$ (including $F > 1$).
\section{The COM-Poisson Distribution}
\label{COM sec}
The Conway-Maxwell-Poisson distribution (COM-Poisson) is a two-variable generalization of the Poisson distribution, first proposed by Conway and Maxwell for application to queuing systems \cite{queuing}. In more recent years, it has garnered attention for its utility in modeling underdispersed and overdispersed data \cite{usefulCOM}. It has found use in marketing, biology, transportation, and a variety of other applications \cite{mainCOM}. The COM-Poisson probability distribution function for a random variable $X$ is defined as \cite{mainCOM}
\begin{equation}
\begin{gathered}
\label{COM}
P(X = x|\lambda,\nu) = \frac{\lambda^x}{\left(x!\right)^{\nu}Z\left(\lambda,\nu\right)} \\ \mathrm{for} \;\; x \in \mathbb{N}_0, \;\; \lambda > 0, \;\; \nu \geq 0,
\end{gathered}
\end{equation}
\noindent where $Z \left( \lambda,\nu \right)$ is a normalizing constant:
\begin{equation}
\label{Z eq}
Z\left(\lambda,\nu\right) = \sum_{s = 0}^{\infty} \frac{\lambda^s}{\left(s!\right)^{\nu}}.
\end{equation}
The parameter $\nu$ controls the dispersion of the distribution. In particular, having $\nu > 1$ will result in underdispersion, and $\nu < 1$ overdispersion. In the special case of $\nu = 1$ the COM-Poisson distribution reduces to the regular Poisson distribution, and $\lambda$ simply becomes the expectation value. The COM-Poisson distribution also reduces to the Bernoulli distribution in the limit $\nu \rightarrow \infty$, as well as the geometric distribution when $\nu = 0$ \cite{usefulCOM}.

While computation of the infinite sum $Z$ may seem unpalatable, in the case of underdispersion the sum converges rapidly and so is simple to calculate. An upper bound on the error from truncating the sum at $k+1$ terms is given by \cite{mainCOM}
\begin{equation}
\frac{\lambda^{k+1}}{\left(k+1\right)!^{\nu} \left(1 - \epsilon_k \right)},
\end{equation}
\noindent where $\epsilon_{k} > \lambda \left(j + 1 \right)^{\nu} \;\; \forall j > k$. The first two central moments of the distribution are given by \cite{usefulCOM}
\begin{equation}
\label{moments orig}
E(X) = \frac{\partial \log{Z\left(\lambda,\nu\right)}}{\partial \log{\lambda}}, \quad \quad Var(X) = \frac{\partial E(X)}{\partial \log{\lambda}}.
\end{equation}
However, this representation does not easily lend itself to computation, so the mean $\mu$ and variance $\sigma_N^2$ can instead be expressed as infinite sums by substituting Eq.\ (\ref{Z eq}) into Eq.\ (\ref{moments orig}):
\begin{equation}
\begin{aligned}
\label{moments sum}
\mu \left( \lambda,\nu \right) &= \sum_{s = 0}^{\infty} \frac{s \lambda^s}{\left(s!\right)^{\nu} Z \left( \lambda, \nu \right)}, \\ \sigma_N^2 \left( \lambda,\nu \right) &= \sum_{s = 0}^{\infty} \frac{s^2 \lambda^s}{\left(s!\right)^{\nu} Z \left( \lambda, \nu \right)} - \mu^2.
\end{aligned}
\end{equation}
\indent As with the normalizing constant $Z$, these sums converge relatively quickly and are easy to compute to arbitrary precision. One useful property of the COM-Poisson distribution is that values of the PDF can be calculated recursively \cite{usefulCOM} using
\begin{equation}
\label{recur eq}
\frac{P\left( X = x - 1 \right)}{P\left( X = x \right)} = \frac{x^\nu}{\lambda}.
\end{equation}
We exploit this for more efficient computation of the CDF of the distribution, which is used for generation of random numbers. A complete overview of the basic properties of the COM-Poisson distribution is given in \cite{usefulCOM,mainCOM,compute}. Notably, much work has been done on fitting data with the COM-Poisson distribution using likelihood, Bayesian, and other techniques \cite{mainCOM,bayesian}.

While the COM-Poisson distribution has many appealing properties, one problem with it for the application of modeling ionization statistics is that the distribution parameters $\lambda$ and $\nu$ do not correspond to the mean, variance, or Fano factor, or indeed to anything with physical meaning. A large part of the present work is dedicated to strategies for overcoming this.
\begin{figure}[!b]
\center
\includegraphics[width=0.445\textwidth]{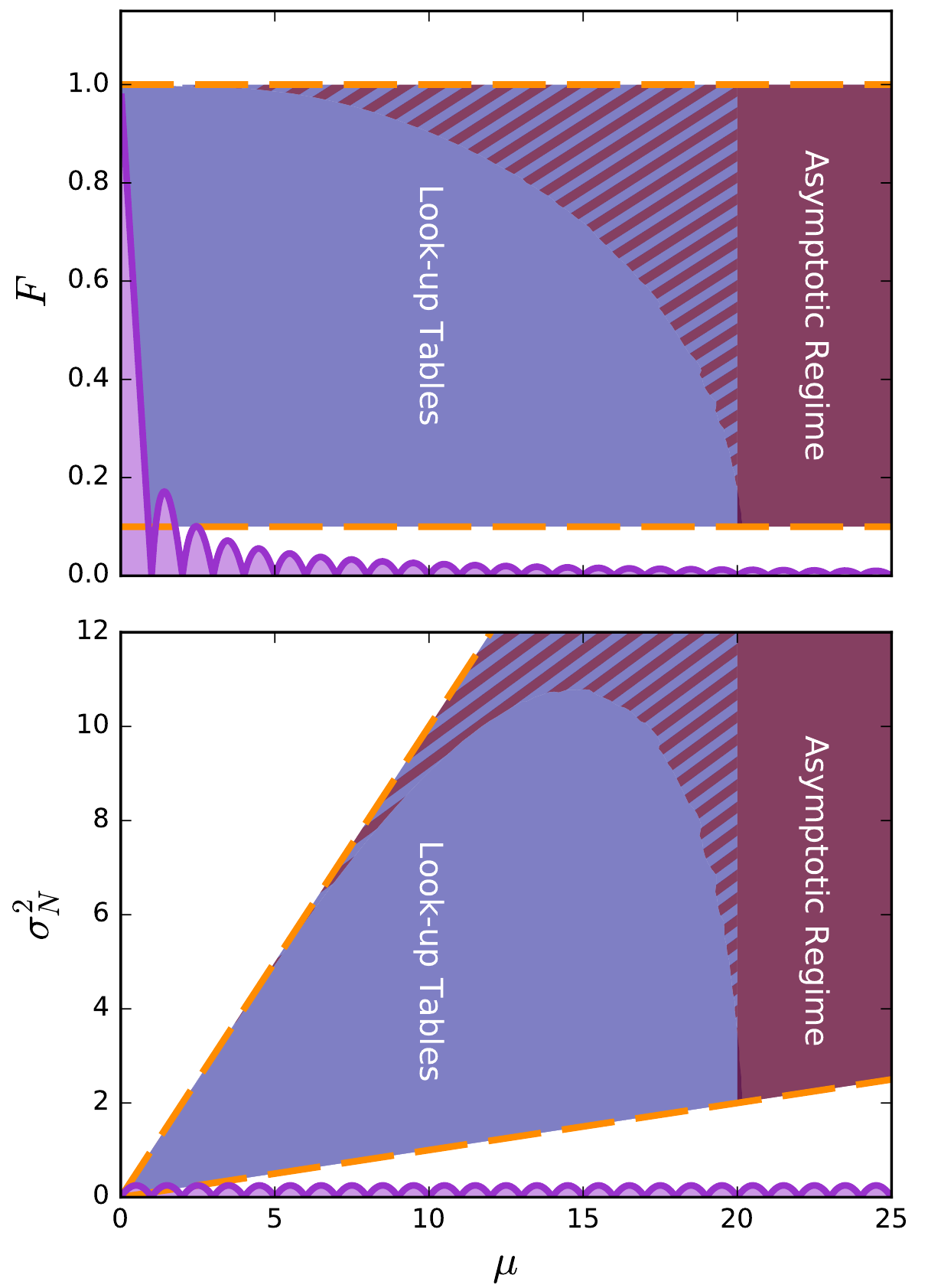}
\caption{Strategies for using the COM-Poisson distribution shown in ($\mu,F$) parameter space (top) and ($\mu,\sigma_N^2$) parameter space (bottom). The blue shaded region represents parameter space where the optimization algorithm and look-up tables will be used (see Sec.\ \ref{optimization sec}). The red shaded region represents the regime where the asymptotic expressions will be used (see Sec.\ \ref{asymp sec}). In the overlap region, the asymptotic expressions are comparably accurate to the optimization algorithm, so points in the look-up tables in this area are calculated with the more accurate of the two. These regions are bounded by the limits of $F$ that we are concerned with (orange dashed lines), namely $0.1 \leq F \leq 1$, and by the Bernoulli modes (purple bumps).}
\label{param fig}
\end{figure}
\section{Using the COM-Poisson Distribution}
\label{using sec}
To model ionization statistics, one needs to be able to independently specify the mean $\mu$ and Fano factor $F$ of the discrete probability distribution function of the number $N$ of pairs created. Because the distribution parameters of COM-Poisson do not correspond to these variables, a method to obtain $\lambda$ and $\nu$ yielding the desired $\mu$ and $F$ is needed. In other words, one needs a mapping between ($\lambda,\nu$) and ($\mu,F$) parameter space to reexpress the COM-Poisson distribution $P(N|\lambda,\nu)$ as $P(N|\mu,F)$. What follows is a discussion of strategies to do so in different regions of ($\mu,F$) parameter space. The three regimes addressed in the subsections below are identified in Fig.\ \ref{param fig}.
\subsection{Asymptotic regime}
\label{asymp sec}
Work has been done to develop an asymptotic expression for $Z$, which to first order can be expressed as \cite{usefulCOM}
\begin{equation}
\begin{aligned}
\label{Z asymp}
Z_{\mathrm{asymp}} \left(\lambda,\nu\right) = &\frac{e^{\nu\lambda^{1/\nu}}}{\lambda^{(\nu-1)/(2\nu)} (2\pi)^{(\nu-1)/2} \sqrt{\nu}} \\ &\times \left(1+\mathcal{O}\left(\lambda^{-1/\nu}\right)\right).
\end{aligned}
\end{equation}
This asymptotic expression is nominally accurate when $\nu < 1$ or when $\lambda > 10^{\nu}$ \cite{usefulCOM}. Beyond making the computation of $Z$ easier, this expression has greater applicability in the context of this work. It can be used to approximate the mean and variance of the distribution for $\lambda$ and $\nu$ by substituting this expression into Eq.\ (\ref{moments orig}). From this we have \cite{usefulCOM,compute}
\begin{equation}
\begin{gathered}
\label{Mean approx}
\mu \approx \lambda^{1/\nu} - \frac{\nu - 1}{2\nu}, \quad \quad \sigma_N^2 \approx \frac{1}{\nu}\lambda^{1/\nu}.
\end{gathered}
\end{equation}
\indent Thus there are closed form expressions for the mean and the Fano factor wherever $Z_{\mathrm{asymp}}$ is accurate. We can now treat these as a system of equations and solve for the distribution parameters as functions of $\mu$ and $F$:
\begin{equation}
\begin{gathered}
\label{nu approx}
\nu(\mu,F) \approx \frac{2\mu + 1 + \sqrt{4\mu^2 + 4\mu + 1 - 8\mu F}}{4\mu F}, \\ \lambda(\mu,F,\nu) \approx \left(\mu \nu F\right)^{\nu}.
\end{gathered}
\end{equation}
\indent These expressions provide a way to calculate $\lambda$ and $\nu$ directly. For the sake of simplicity, we choose to utilize them above $\mu = 20$ (the red shaded region in Fig.\ \ref{param fig}) where we have verified that they are accurate to $0.01\%$ or better for both $\mu$ and $F$. The asymptotic approximation for $Z$ solves several other problems as well. As mentioned in Sec.\ \ref{COM sec}, $Z$ converges quickly in the case of underdispersion ($\nu > 1$), but not so in the case of overdispersion \cite{mainCOM}, so it may be necessary to use Eq.\ (\ref{Z asymp}) for $\nu < 1$. Another computational challenge with $Z$ arises for underdispersion when $\lambda$ becomes very large, at which point $Z$ itself becomes so large that it is not storable as a normal double-precision value. In this case, it is necessary to calculate $\log{Z}$ directly using Eq.\ (\ref{Z asymp}).
\subsection{Bernoulli modes}
\label{bernoulli sec}
Because the COM-Poisson distribution is discrete, there is a fundamental lower limit to the variance achievable with it, which is a function of the mean. To see how this is the case, consider a situation in which one has integer data with a mean of $0.5$ and a variance of $0.25$. This necessarily means that there are an equal number of counts of $0$ and $1$. In fact, it is impossible for the data to have a smaller variance, as this would require more counts to be either $0$ or $1$, which would shift the mean of the data. A slightly smaller or larger mean allows for a smaller variance, but there is still a minimum variance given by $\mu (1-\mu)$. This is the regime where discrete probability distributions reduce to the Bernoulli distribution, which the COM-Poisson distribution does when $\nu \rightarrow \infty$. 

This also applies to larger values of the mean, creating an unending series of ``Bernoulli modes" shown in Fig.\ \ref{param fig} defining the minimum possible variance of discrete data, which are given by
\begin{equation}
\sigma^2_{N_{min}} = -(\mu - \mu_i)(\mu - \mu_i - 1)
\end{equation}
\noindent with $\mu_i \in \mathbb{N}$ such that $\mu_i - 1 < \mu < \mu_i$. This means that some ($\mu,F$) parameter space is fundamentally inaccessible by the COM-Poisson distribution or any other discrete distribution. As a consequence of this, the Fano factor of any material necessarily tends to $1$ when $\mu \ll 1$. Using the COM-Poisson distribution near the Bernoulli modes is also difficult as $\nu \rightarrow \infty$. To avoid this issue, we transition to using the Bernoulli distribution when within a distance of $0.1\%$ in ($\mu,F$) parameter space of the Bernoulli modes.
\subsection{Optimization algorithm}
\label{optimization sec}
\begin{figure*}[t]
\subfigure{\includegraphics[width=0.32\textwidth]{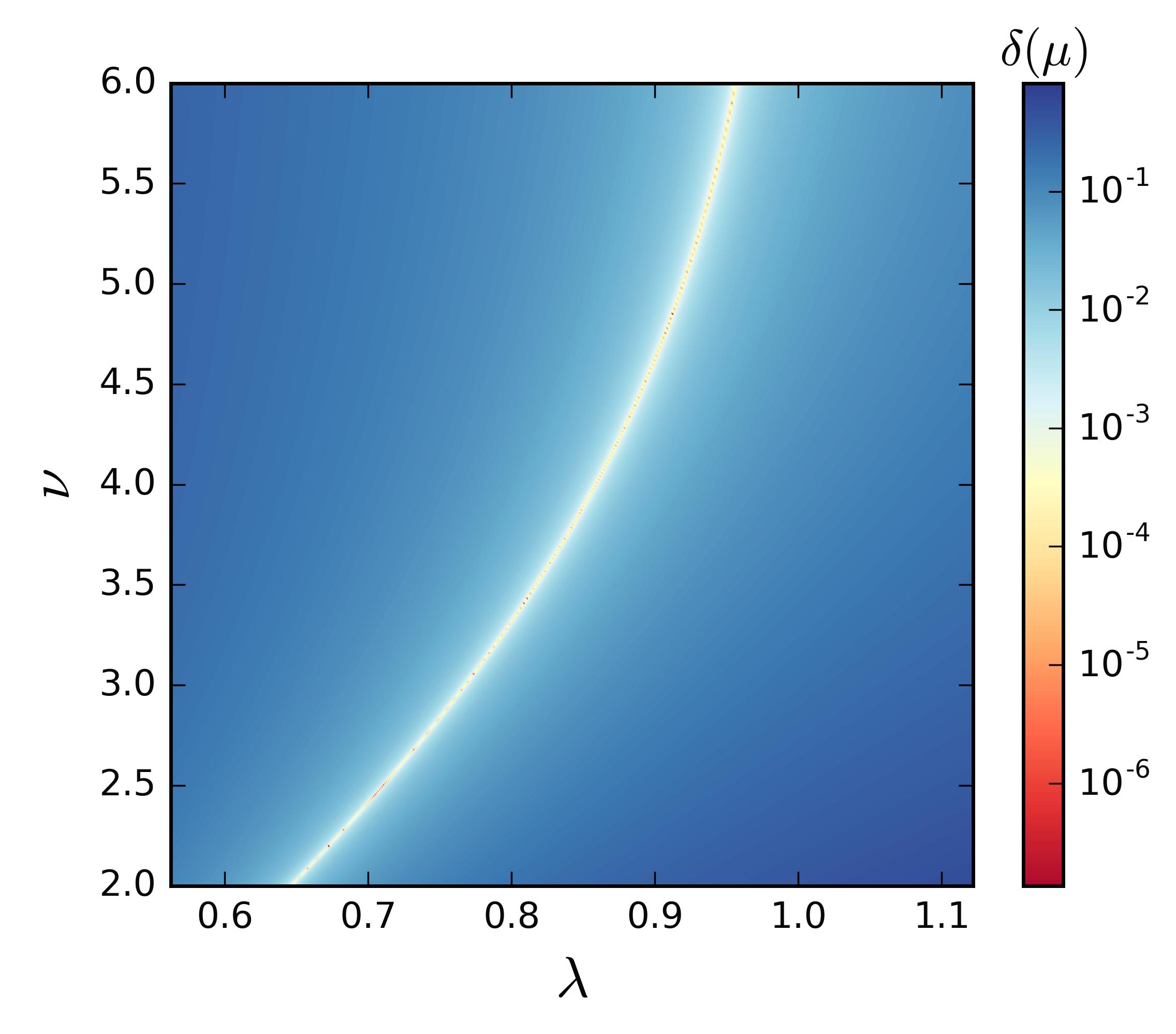}}
\subfigure{\includegraphics[width=0.32\textwidth]{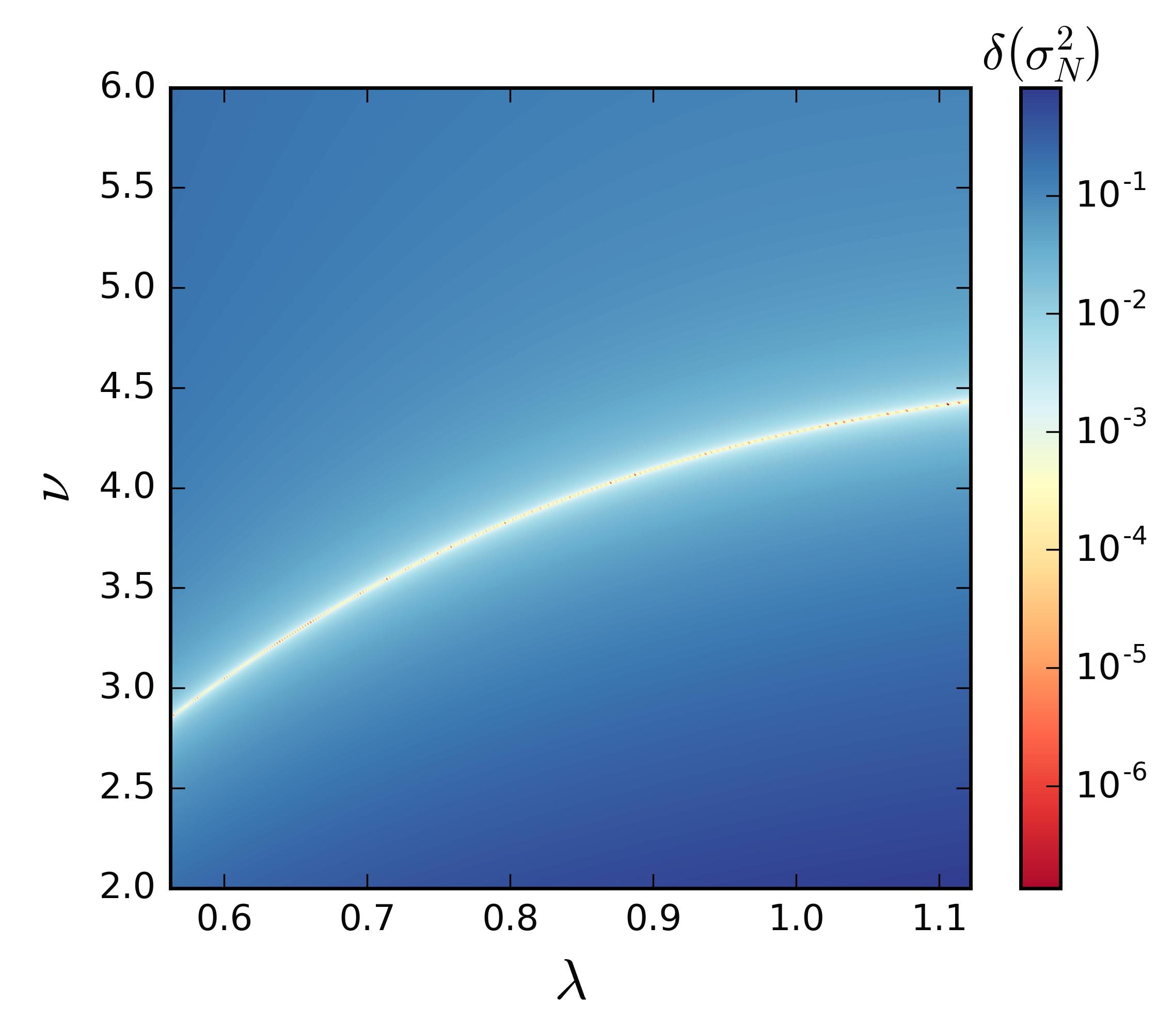}}
\subfigure{\includegraphics[width=0.32\textwidth]{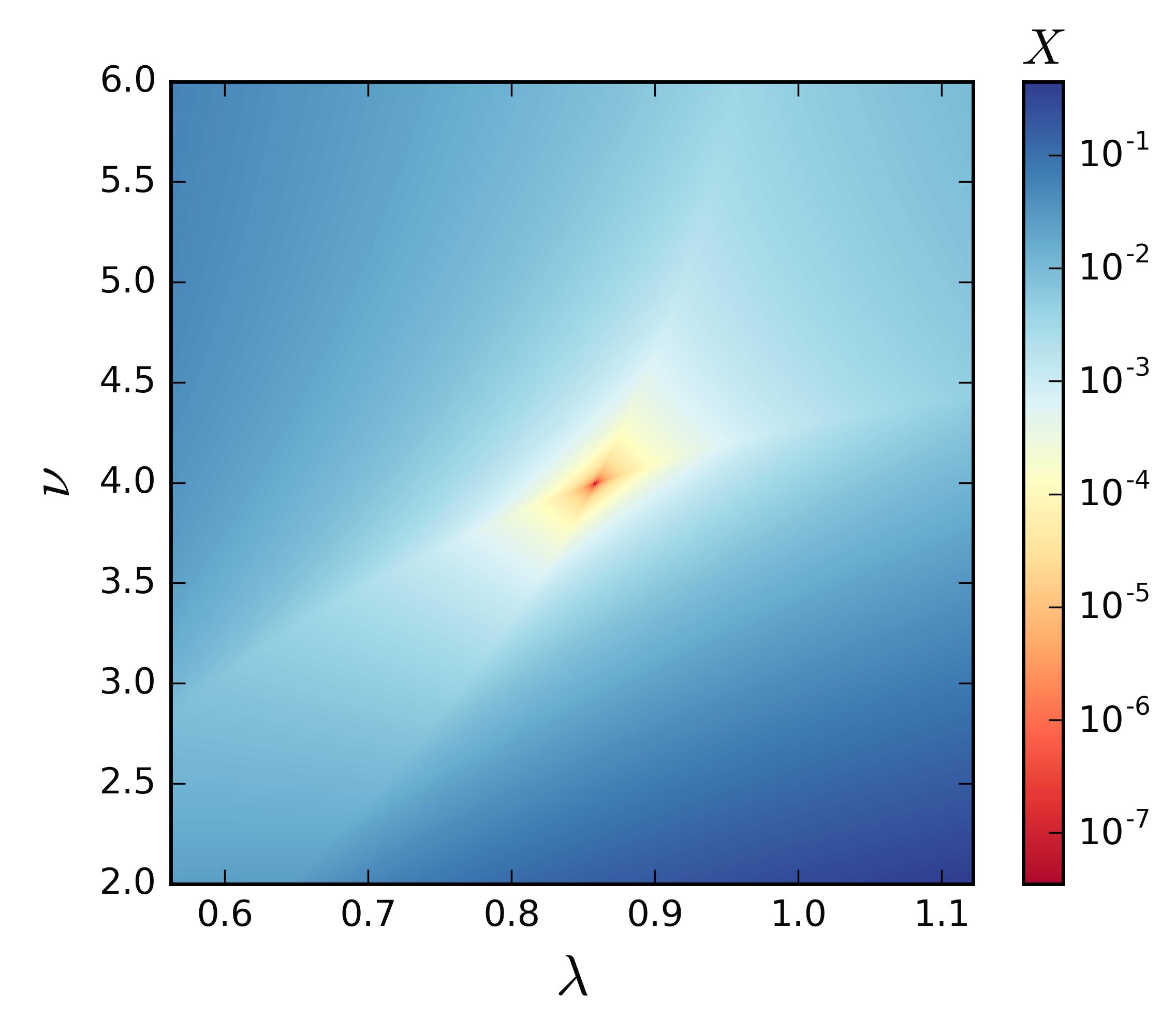}}
\caption{Illustrative example of the minimization algorithm for $\mu = 0.5$ and $F = 0.6$. First (left) the relative error in the mean $\delta(\mu)$ is shown as a function of $\lambda$ and $\nu$, and then (middle) the relative error in the variance $\delta \left(\sigma_N^2 \right)$. Both plots clearly show a ``valley" where the desired mean and variance are obtainable. Then (right) the weighted average of the relative error in mean and variance (the quantity $X$) is plotted as a function of $\lambda$ and $\nu$, effectively overlapping the two ``valleys." Thus, the values of $\lambda$ and $\nu$ giving the desired $\mu$ and $F$ can be found by minimizing $X$.}
\label{op fig}
\end{figure*}
At this point, we have addressed a large portion of ($\mu,F$) parameter space. At low values of $F$, the Bernoulli modes provide a lower bound on the accessible parameter space. We know empirically that $F \leq 1$ and that $F \geq 0.1$ for most detectors. At $\mu \gtrsim 20$ we have asymptotic expressions for $\lambda$ and $\nu$ as functions of $\mu$ and $F$ that allow us to control the COM-Poisson distribution directly. At even higher values of $\mu$ there is also the option to use a Gaussian distribution to model ionization statistics to save computation time.

\begin{figure}[b!]
\center
\includegraphics[width=0.41\textwidth]{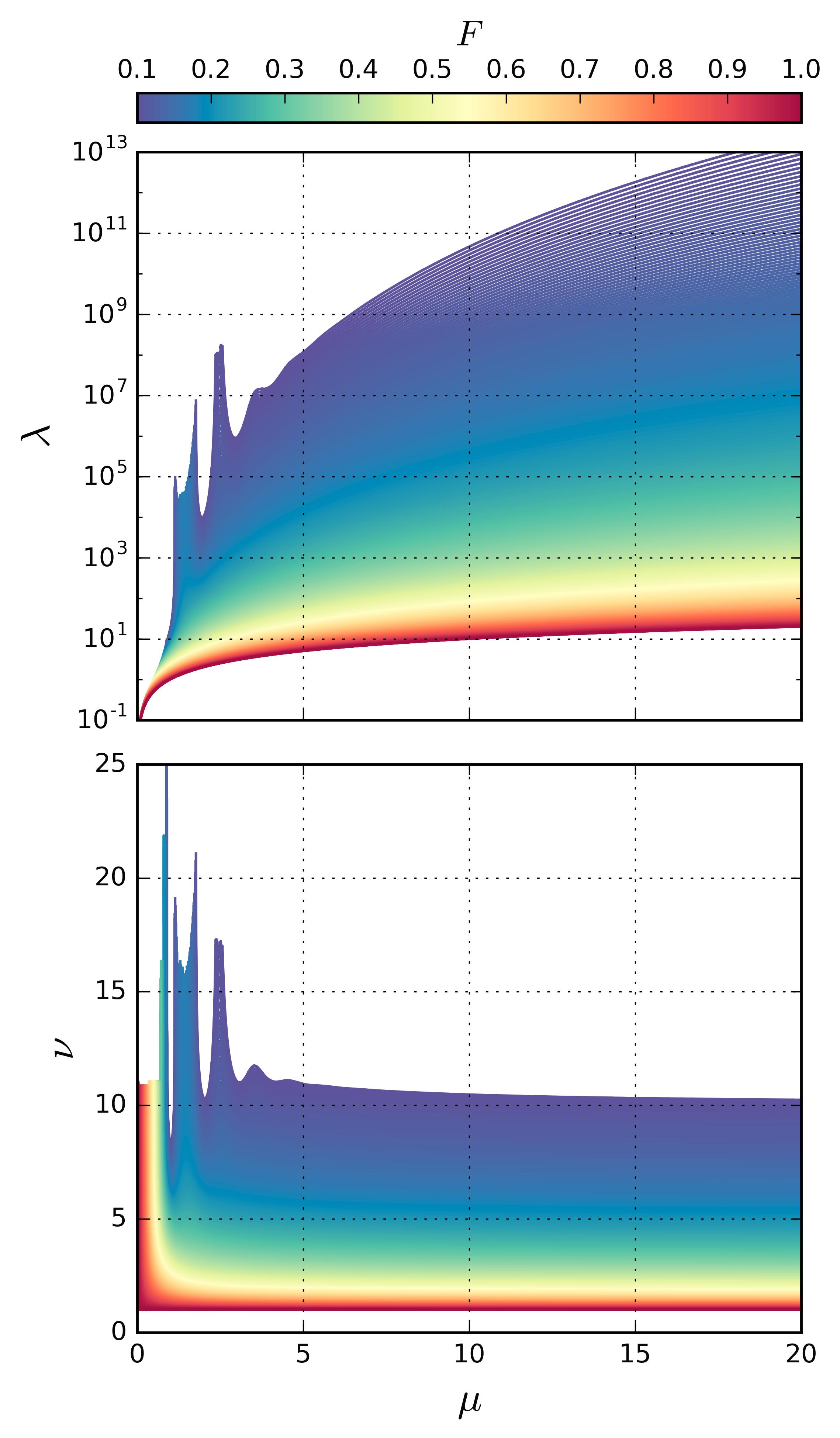}
\caption{The COM-Poisson distribution parameters $\lambda$ and $\nu$ as a function of $\mu$ and $F$ (color scale) as obtained by the optimization algorithm, which are contained in the look-up tables.}
\label{rainbow fig}
\end{figure}

However, there is still a large region of parameter space yet to be addressed between the Bernoulli modes and the asymptotic regime, shown in blue in Fig.\ \ref{param fig}. Thus another strategy to determine the distribution parameters $\lambda$ and $\nu$ for a desired $\mu$ and $F$ is necessary. The approach we employed was to use an optimization algorithm. We show an illustrative example of this method in Fig.\ \ref{op fig}. For given values of $\lambda_i$ and $\nu_i$ we calculated the relative error between the desired mean and variance $\left( \mu \;\mathrm{and} \; \sigma_N^2\right)$ and obtained mean and variance $\left(\mu_i \; \mathrm{and} \; {\sigma_N^2}_{i} \right)$ calculated with Eq.\ (\ref{moments sum}). In both cases there is a valley of ($\lambda,\nu$) parameter space giving the desired mean and variance. To reduce this problem to a single scalar minimization problem, and to obtain the unique $\lambda$ and $\nu$ that give us the mean and variance we want, we define the following quantity $X$:
\begin{equation}
X\left(\lambda,\nu\right | \mu,\sigma_N^2) = \left( w_1 \left| \frac{\mu - \mu_i}{\mu} \right| + w_2 \left| \frac{\sigma_N^2 - {\sigma_N^2}_{i}}{\sigma_N^2} \right| \right)^p.
\end{equation}
\indent This is the weighted average of the relative error in mean and variance for a given $\lambda$ and $\nu$. This quantity is then minimized to find the desired values of $\lambda$ and $\nu$. The weights $w_1$ and $w_2$ and exponent $p$ were tuned to improve the performance of the optimization. More details about this algorithm are presented in the Appendix. While effective, this algorithm is not always practical to use because of the required computation time, taking on average one second per execution, and potentially longer if convergence is not obtained easily. For applications where the distribution must be used thousands of times with different $\mu$ and $F$ each time, this is not practical.
\indent To overcome this, we have executed this optimization algorithm for a dense grid of points in $\mu$ and $F$ and saved the results in ``look-up tables" of values for $\lambda$ and $\nu$. The tables span the region $0 < \mu < 20$ and $0.1 < F < 1$, and the values in them are accurate to $0.1\%$ in $\mu$ and $F$ or better. The full contents of the tables are presented in Fig.\ \ref{rainbow fig}. Interpolation between the points is also possible, so that the COM-Poisson distribution can be used for any value of $\mu$ and $F$ within the scope of the tables. Details about this and how the accuracy of the tables is determined are also given in the Appendix and \cite{website}.
These look-up tables along with the asymptotic expressions constitute a comprehensive, practical strategy for using the COM-Poisson distribution for any desired $\mu$ and $F$. This makes it a viable tool for modeling ionization statistics in particle detectors.
\section{Applications}
\label{App sec}
\begin{figure}[htp!]
\center
\includegraphics[height=7cm,keepaspectratio]{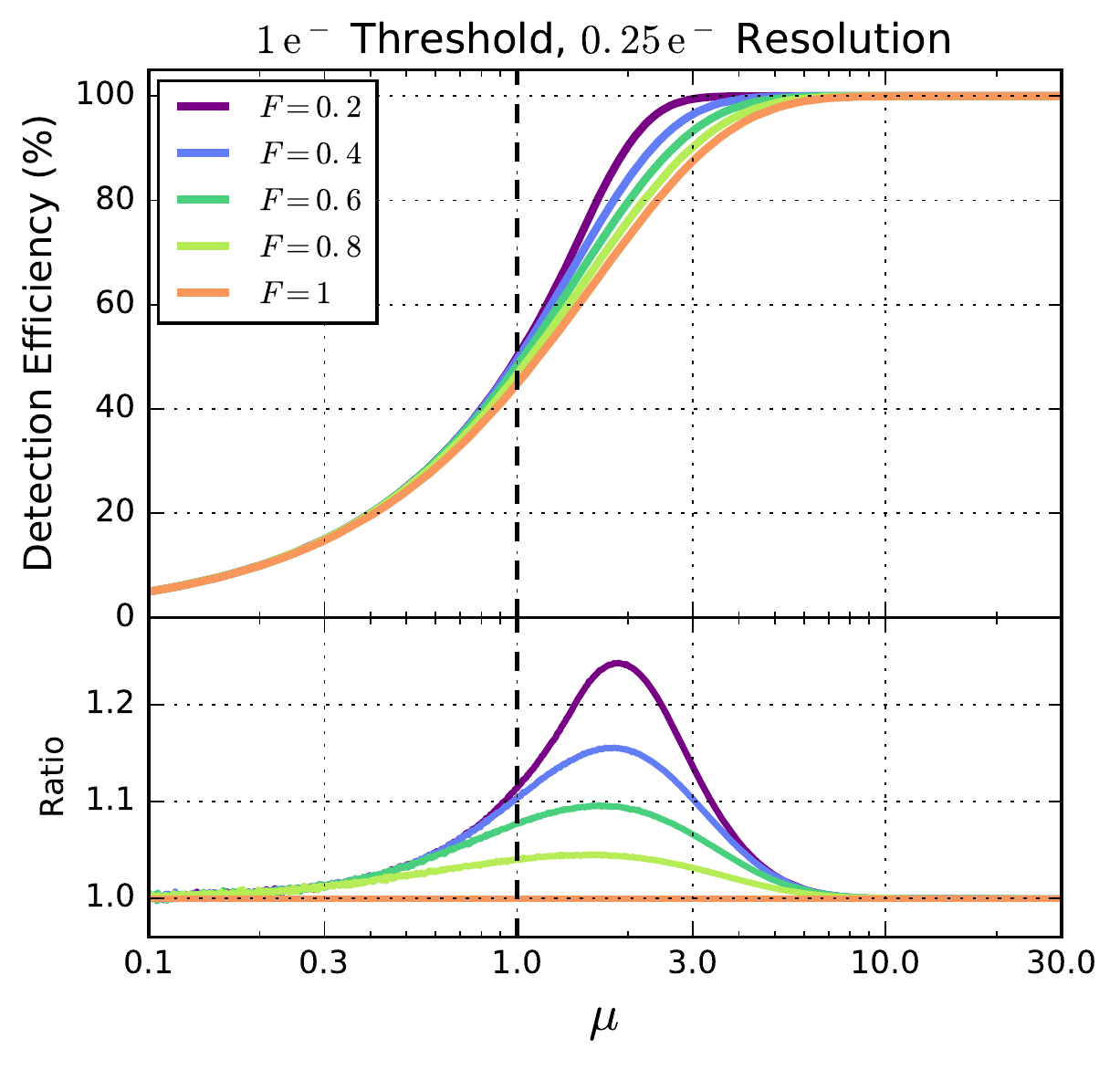}
\includegraphics[height=7cm,keepaspectratio]{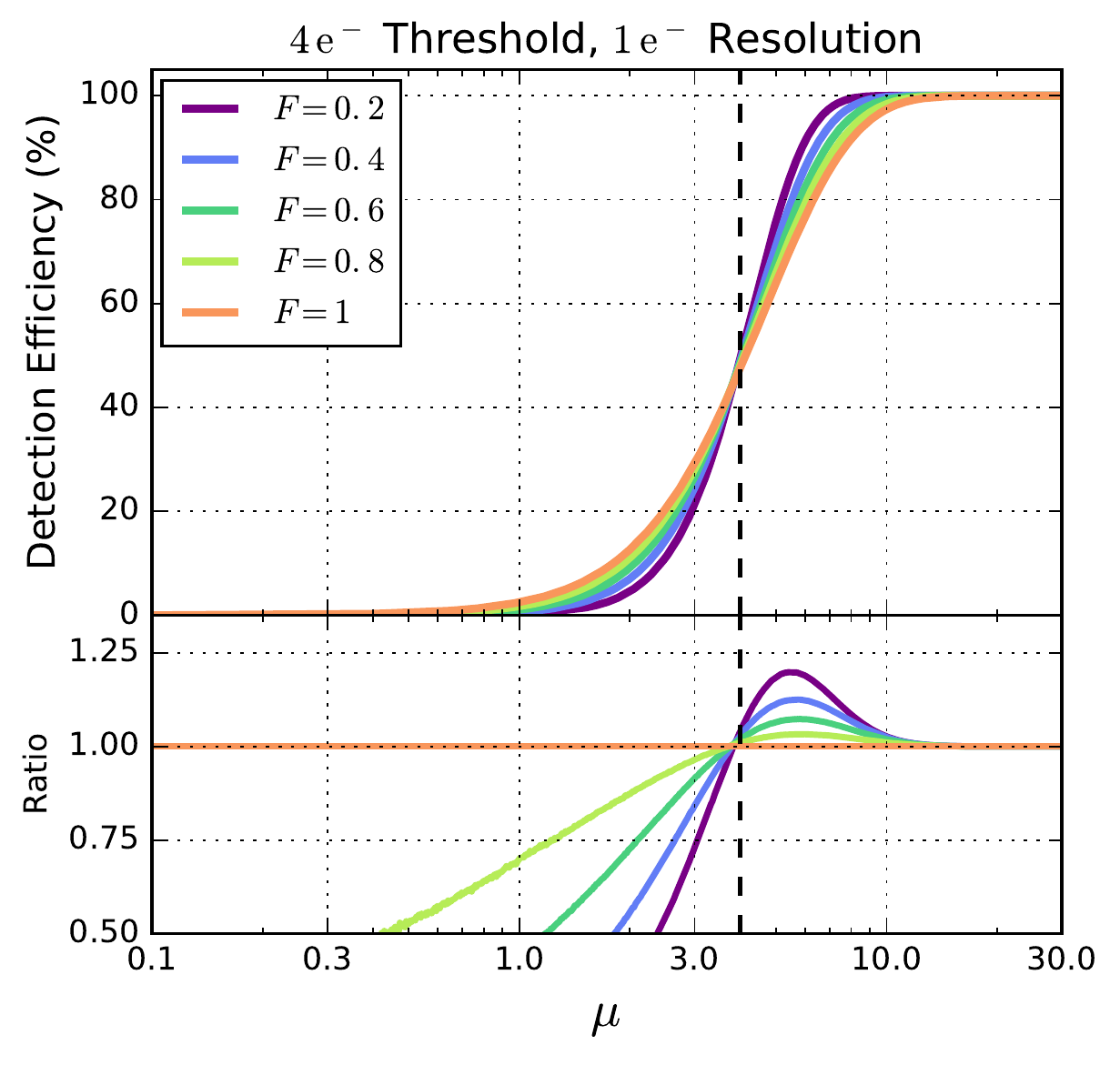}
\includegraphics[height=7cm,keepaspectratio]{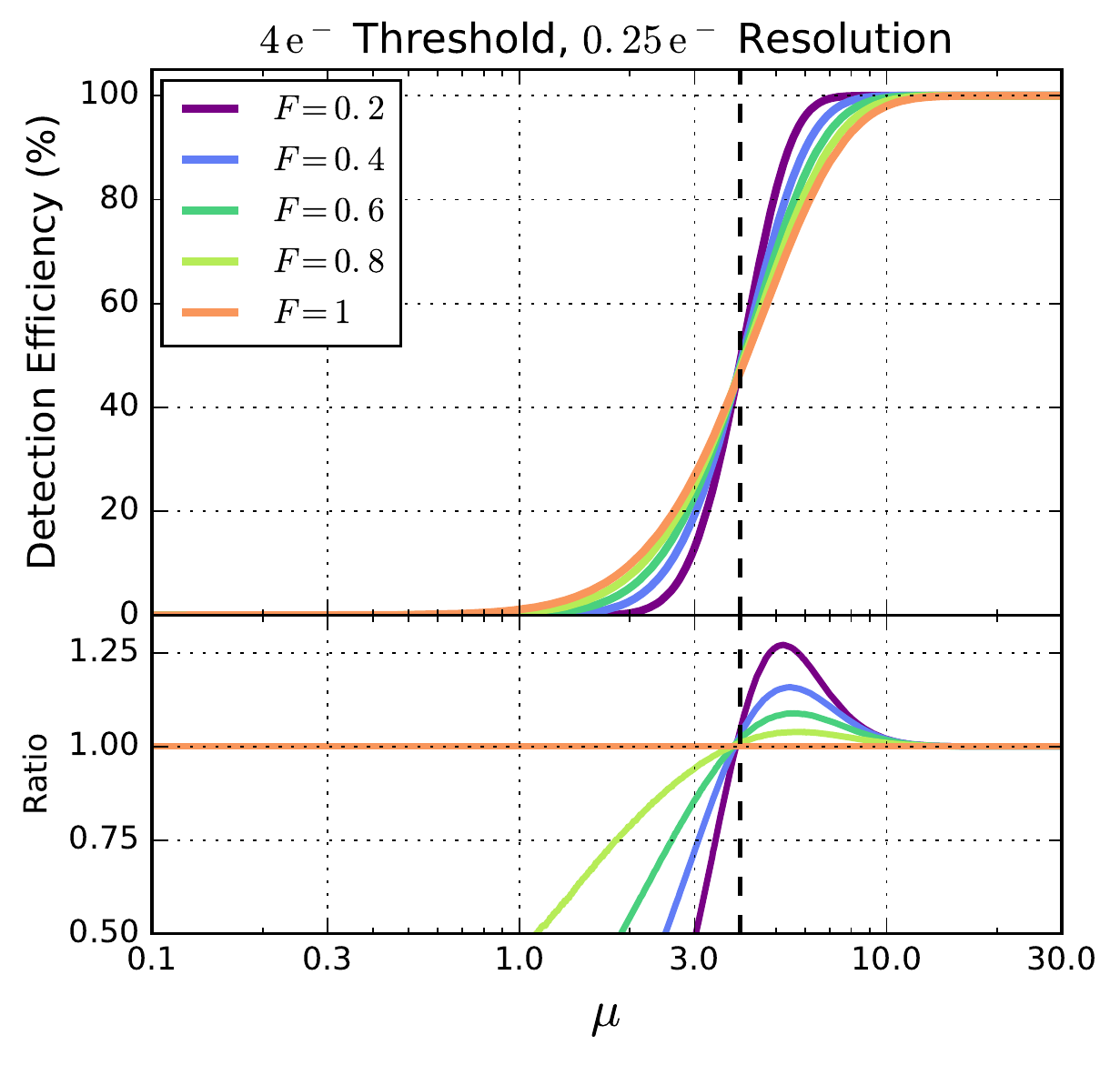}
\caption{Detection efficiency of an experiment with a $1\,\mathrm{e^-}$ (top) and a $4\,\mathrm{e^-}$ (middle and bottom) energy threshold, using the COM-Poisson distribution to model ionization statistics and with $\sigma = 0.25\,\mathrm{e^-}$ (top and bottom) and $\sigma = 1\,\mathrm{e^-}$ (middle) Gaussian resolution, with given values of $F$ (where possible, see Sec.\ \ref{bernoulli sec}). The ratio of the curves with respect to $F=1$ are shown in the subpanels.}
\label{Survive fig}
\end{figure}

We now make use of this tool to assess how ionization statistics affects the sensitivity of particle detectors to low energy events. In order to remain as general as possible, we first study the impact of the Fano factor on an experiment's detection efficiency. This approach has the advantage of not needing to make any assumptions about the spectral shape of the signal being observed. In Fig.\ \ref{Survive fig} we show detection efficiency curves obtained from Monte Carlo simulations where for each given expected number of pairs $\mu$, the actual number of pairs created is randomly drawn $10^6$ times from the COM-Poisson distribution with a fixed value of $F$. The detection efficiency is then calculated as the proportion of events reconstructed above threshold and is shown as a function of energy in units of $\mu$. To assess how energy threshold and resolution (e.g. baseline fluctuations) may change the impact of the Fano factor, Fig.\ \ref{Survive fig} shows  detection efficiency curves computed for three different scenarios of threshold $E_{\mathrm{th}}$ and detector resolution, which is modeled with a Gaussian distribution of standard deviation $\sigma$. The three  scenarios considered are $E_{\mathrm{th}} = 1\,\mathrm{e^-}$/$\sigma = 0.25\,\mathrm{e^-}$ (top panel), $E_{\mathrm{th}} = 4\,\mathrm{e^-}$/$\sigma = 1\,\mathrm{e^-}$ (middle panel), and $E_{\mathrm{th}} = 1\,\mathrm{e^-}$/$\sigma = 0.25\,\mathrm{e^-}$ (bottom panel). For each, the ratio with respect to $F=1$ is shown in the subpanel.

From these results we can observe that in all scenarios, lower values of $F$ increase sensitivity to events with $\mu > E_{\mathrm{th}}$, as this results in a reduced probability of these being reconstructed below threshold. For events with $\mu < E_{\mathrm{th}}$, one expects the effect to be reversed due to low values of $F$ decreasing the probability of an event being reconstructed above threshold, and this is indeed observed in the $4\,\mathrm{e^-}$ threshold scenarios (lower panels). However, the trend is nontrivially different in the $1\,\mathrm{e^-}$ threshold scenario in which all the efficiency curves quickly tend to overlap in the low energy regime. This should not be misinterpreted as the Fano factor having no impact on the detection efficiency, but rather that when $\mu \ll 1$ the Fano factor is naturally forced to converge to $1$ as the Bernoulli regime is reached (see Sec.\ \ref{bernoulli sec}). One can also see that an improvement of the resolution from $\sigma = 1\,\mathrm{e^-}$ to $\sigma = 0.25\,\mathrm{e^-}$ tends to magnify the impact of the Fano factor over the whole energy range. Ultimately, an extremely poor resolution would make the detection efficiency almost insensitive to the Fano factor as resolution effects would dominate over fluctuations in pair creation. Finally, one can conclude from the lower panels of Fig.\ \ref{Survive fig} that the impact of $F$ can be most extreme for the detection of events below a high energy threshold.

From the above, we can infer that the ramifications of the Fano factor will depend on how crucial subthreshold event detection is to an experiment. For direct dark matter detection experiments searching for low-mass weakly interacting massive particles (WIMPs), this is all the more true. They are faced with the challenge of measuring the minute recoil energies of target nuclei following a WIMP elastic scattering interaction. The theoretical recoil energy spectrum of these events is approximately an exponential distribution with a slope that increases and a maximum energy cutoff that decreases as the WIMP mass decreases \cite{lewin, schnee}. For very low WIMP masses, an experiment's sensitivity may derive primarily, if not entirely, from subthreshold events. An in-depth study of how the Fano factor may affect specific, existing experiments is out of the scope of this paper. Indeed, this could strongly depend on numerous factors which vary from one experiment to another including the target atomic number A, quenching for that target (ionization yield), and energy resolution. Rather, we wish to demonstrate the importance of accounting for the Fano factor when deriving sensitivity to low-mass WIMPs and to show that the COM-Poisson distribution proves to be a useful tool to do so. To that end, we considered a hypothetical WIMP search experiment with a target arbitrarily chosen to be neon and the quenching parametrization used by \cite{Arnaud2018}. We also incorporate the most generic detector energy resolution by modeling it with a Gaussian distribution. We use the WIMP recoil energy spectrum given by \cite{lewin} with a local dark matter density of $\rho_{\chi} = 0.3 \, \mathrm{GeV/cm^3}$ and standard halo parameters.

\begin{figure}[htp!]
\center
\includegraphics[height=7cm,keepaspectratio]{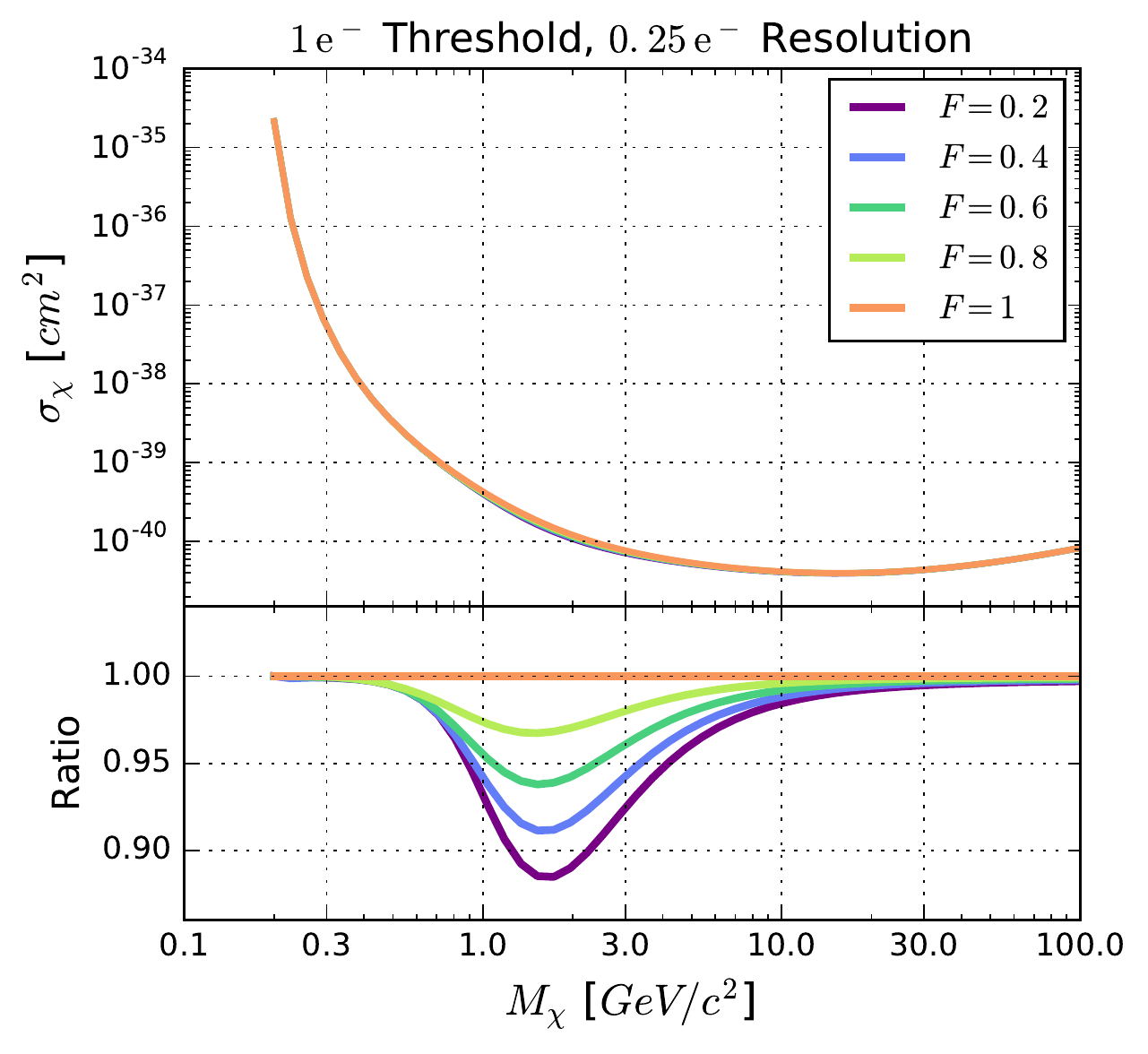}
\includegraphics[height=7cm,keepaspectratio]{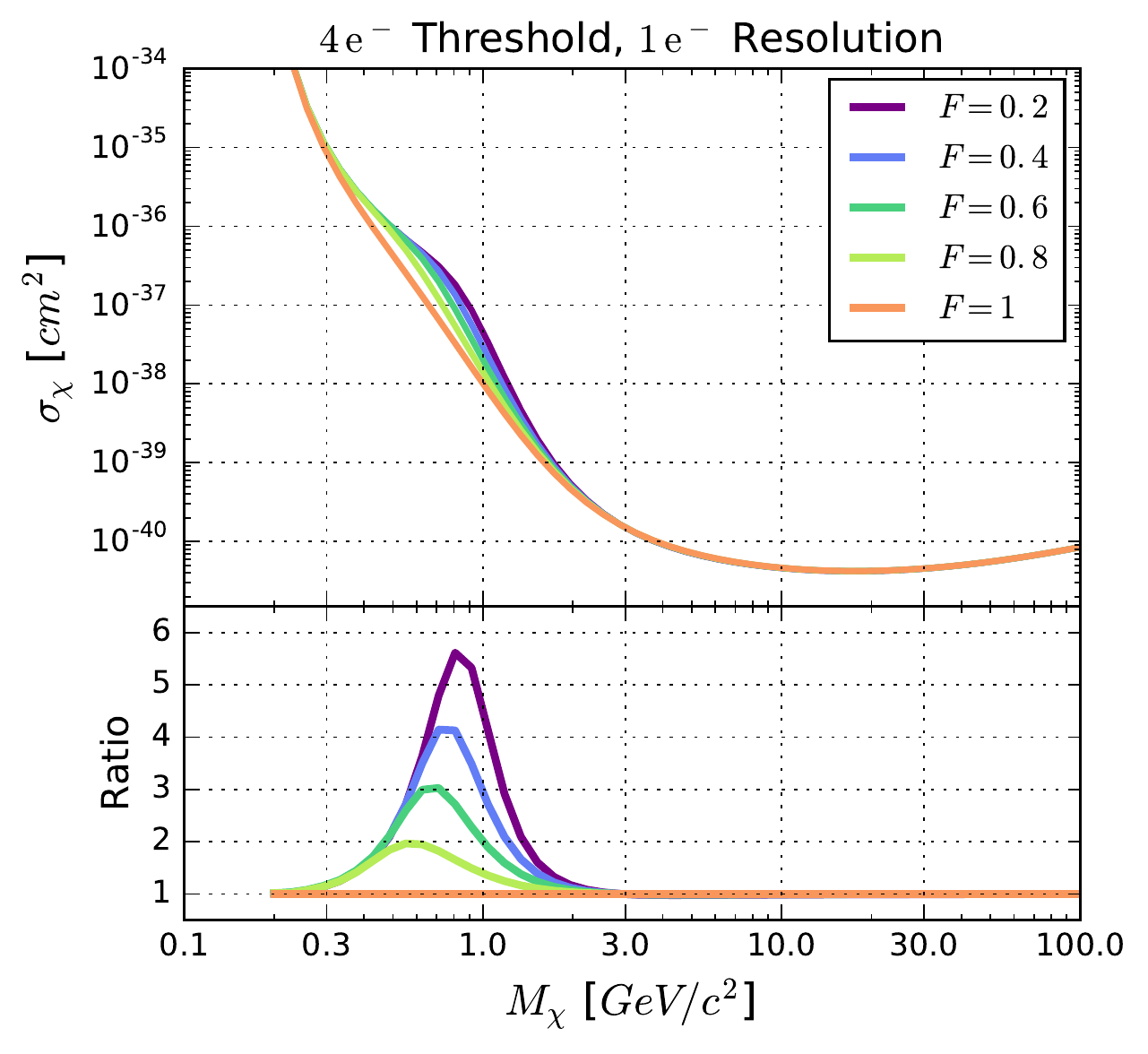}
\includegraphics[width=7.647058824cm,keepaspectratio]{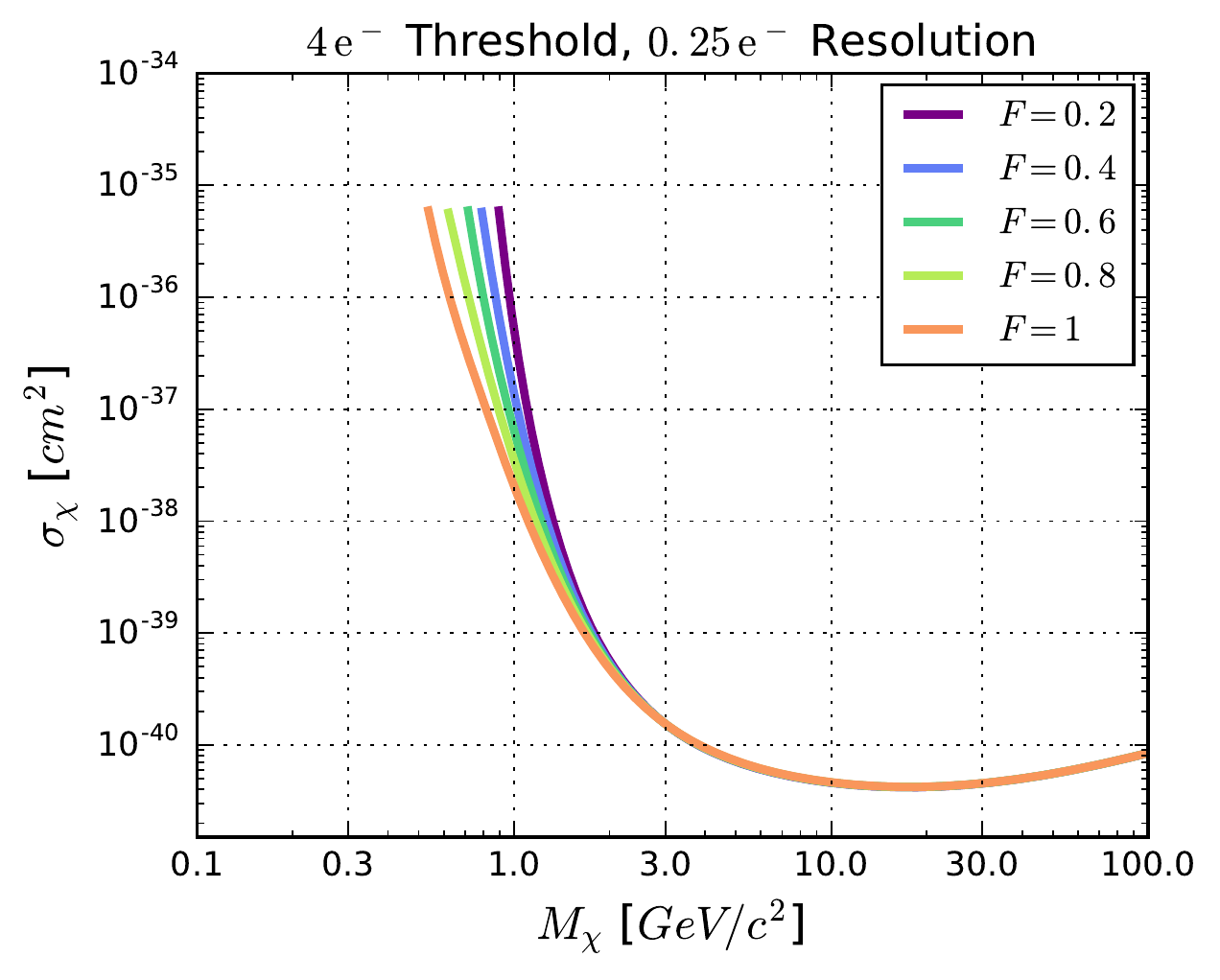}
\caption{Spin-independent WIMP-nucleon scattering cross section exclusion curves for a hypothetical experiment with a neon target, assuming a $1\,\mathrm{e^-}$ (top) and a $4\,\mathrm{e^-}$ (middle and bottom) energy threshold. The energy resolution of the experiment is simulated with the COM-Poisson distribution and Gaussian resolution with $\sigma = 0.25\,\mathrm{e^-}$ (top and bottom) and $\sigma = 1\,\mathrm{e^-}$ (middle). The ratios of the curves with respect to $F=1$ are plotted in the subpanels where appropriate.}
\label{Limit fig}
\end{figure}

We show in Fig.\ \ref{Limit fig} the background-free sensitivities derived for different values of the Fano factor in the same three threshold/resolution scenarios depicted in Fig.\ \ref{Survive fig}. These curves correspond to the $90\%$ confidence level upper limits on the spin-independent WIMP-nucleon elastic scattering cross section ($\sigma_{\chi}$) an experiment would report in the absence of any signal, as a function of WIMP mass $M_{\chi}$. These were calculated assuming an arbitrary exposure of $1 \,\mathrm{kg \cdot day}$, although background-free limits simply scale with exposure. In the first two cases, $F$ only has a significant impact at intermediate WIMP masses. At high masses, WIMP sensitivity primarily comes from events with energies far above threshold where the detection efficiency of the experiment is $100\%$ regardless of $F$ as long as no upper analysis threshold is set. At low WIMP masses, sensitivity is dominated by Bernoulli-regime events with $\mu < 1$ where $F$ is naturally bounded. However, in the low threshold case (top), a smaller value of $F$ conveys a greater sensitivity (albeit only slightly) to $0.5-10\,\mathrm{GeV/c^2}$ WIMPs, as the detection efficiency is always higher for lower $F$ (see Fig.\ \ref{Survive fig} top). In the scenario of high threshold but poor energy resolution (middle), the effect of $F$ at intermediate masses is reversed and far greater in magnitude. This can be understood by considering that in the $0.3-2\,\mathrm{GeV/c^2}$ mass range, most WIMP scattering events are below threshold, and so a larger value of $F$ can dramatically increase sensitivity (see Fig.\ \ref{Survive fig} middle).

Finally, the bottom panel of Fig.\ \ref{Limit fig} depicts an extreme scenario, combining a high threshold with good resolution. As with the previous scenarios, $F$ has essentially no effect at high mass. However, in this case the limits do not converge at low WIMP masses because the experiment is simply not sensitive to Bernoulli-regime events at all. For this reason the impact of the Fano factor at low masses is extreme, with small values of $F$ inducing a multiple order of magnitude reduction in sensitivity. Ultimately, at the lowest WIMP masses an experiment may not be sensitive at all depending on the value of $F$.

\clearpage

\clearpage

\section{Discussion and Conclusions}
\label{discussion sec}
In this work, we have proposed a novel approach and developed the strategies and tools to account for the Fano factor at the level of pair creation with the COM-Poisson distribution. By using it to assess the impact that the Fano factor may have on the sensitivity of low-mass WIMP search experiments, we have demonstrated both the need for modeling ionization statistics and the usefulness of the COM-Poisson distribution to do so. Although there is no physical motivation for the choice of COM-Poisson other than its apparent suitability for this application, we would like to stress that there is no comparable alternative to the proposed approach at the time of this publication. Additional rationale for using the COM-Poisson distribution can be found in \cite{supp}. To encourage and facilitate the use of this tool by others, we have provided free access to the look-up tables discussed in Sec.\ \ref{optimization sec} and the code to use them at \cite{website}. We have also provided code to produce the detection efficiency curves shown in Fig.\ \ref{Survive fig} as an illustrative example. The authors are open to providing assistance in using the tools developed and discussed in this work.
\section*{ACKNOWLEDGEMENTS}
This research was undertaken, in part, thanks to funding from the Canada Excellence Research Chairs Program. The present work was performed within the NEWS-G Collaboration, and benefited from the input and feedback of its members.
\section*{APPENDIX: DETAILS OF OPTIMIZATION ALGORITHM AND LOOK-UP TABLES}
\label{appendix}
The optimization algorithm described in Sec.\ \ref{optimization sec} uses a combination of a built-in minimizer of ROOT as well as a grid search. This is necessary because the ROOT minimizer was often unable to find the minima of $X$ on its own (see Sec.\ \ref{optimization sec}). First, for a desired $\mu$ and $F$, a box in ($\log_{10}{\lambda},\nu$) parameter space is defined based on the asymptotic approximations for $\lambda$ and $\nu$ [Eq.\ (\ref{nu approx})]. Even at low values of $\mu$ and $F$ where these expressions are not very accurate, they can still serve as a reasonable initial starting point. Then, a ``grid search" is performed in which the value of $X$ is computed (see Sec.\ \ref{optimization sec}) for every point on a $100 \times 100$ grid of points in that box, and the grid point with the lowest value of $X$ is found. Next, another smaller box is set based on this point, and the minimum grid square is used as an initial guess for the ROOT Minuit2 algorithm. Specifically, a bound minimization is carried out with the ``combined" method within the defined box (again, with $\lambda$ on a $\log_{10}$ scale). With this new minima (giving values of $\lambda$ and $\nu$), the $\mu$ and $F$ obtained are calculated with Eq.\ (\ref{moments sum}). If these values yield the desired $\mu$ and $F$ within acceptable tolerance ($0.1\%$ in this case), then the algorithm terminates. If this is not the case, the process can be repeated iteratively within smaller and smaller sections of parameter space with randomly perturbed initial guesses and bounds for the box. These extra measures are rarely necessary, as the first efforts of the algorithm are usually sufficient. However, near the Bernoulli modes the values of $\lambda$ and $\nu$ can vary wildly (see Fig.\ \ref{rainbow fig}), so difficulties do occur.

\begin{figure}[!ht]
\center
\includegraphics[width=0.45\textwidth]{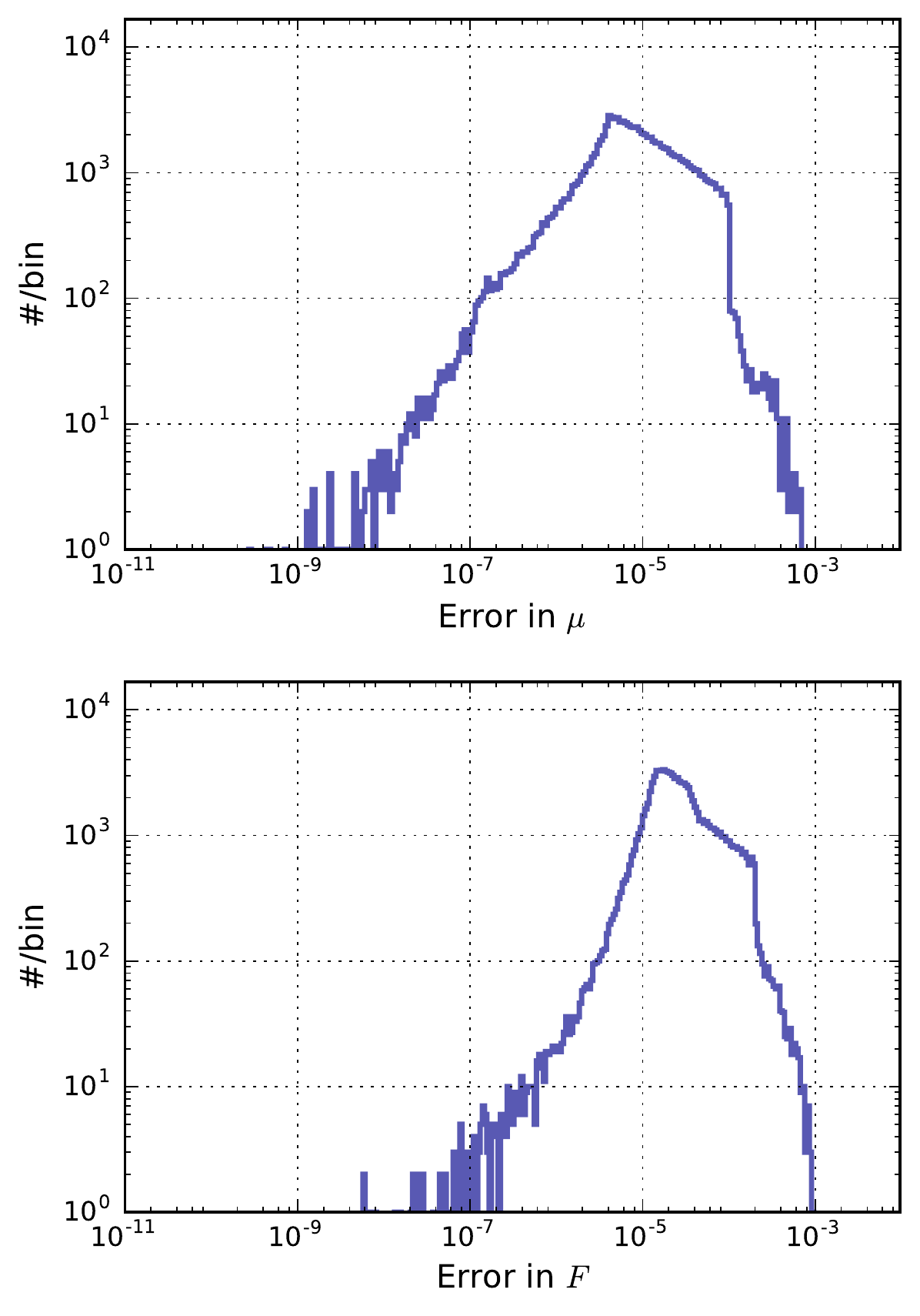}
\caption{Histograms of the error in $\mu$ and $F$ obtained with the look-up tables and asymptotic approximations for $10^6$ points randomly drawn in ($\mu,F$) parameter space within $0 < \mu < 100$ and $0.1 < F < 1$.}
\label{hist fig}
\end{figure}

The version of our look-up tables used for this paper was made with $10,000$ logarithmically spaced values of $\mu$ from $0.001$ to $20$, and $1,000$ linearly spaced values of $F$ from $0.1$ to $1$. Points that fall within $0.1\%$ of Bernoulli modes are excluded. It should be noted that in this scheme, a large proportion of the points are skipped because of this criteria. However, it was important to maintain a greater density of points at lower values of $\mu$ where $\lambda$ and $\nu$ vary more rapidly, as well as maintaining a regular grid to make interpolation possible. The contents of the look-up table are shown in Fig.\ \ref{rainbow fig}, where one can see that the values of $\lambda$ and $\nu$ vary considerably and have emergent features at low values of $F$, defying attempts to conveniently parametrize them. As mentioned in Sec.\ \ref{optimization sec}, ultimately this table should work not just for values of $\mu$ and $F$ directly contained in the table, but for any values through interpolation. To do this, simple code to perform a bilinear interpolation of $\log_{10}{\lambda}/\nu$ was written.

To assess this and ultimately the accuracy of the look-up tables and the asymptotic approximations a test was devised in which random points in $\mu/F$ were chosen within $0 < \mu < 100$ and $0.1 < F < 1$. The appropriate method [either look-up table with interpolation or Eq.\ (\ref{nu approx})] were then used to calculate $\lambda$ and $\nu$ corresponding to the desired values of $\mu$ and $F$, and the error of the obtained vs desired $\mu$ and $F$ was calculated with Eq.\ (\ref{moments sum}). The results of this test for $10^6$ points is shown in Fig.\ \ref{hist fig}. The maximum error obtained in $\mu$ or $F$ for any point was less than or equal to $0.1\%$, as desired, and often much smaller than this. More information about the accuracy of the look-up tables (and code to test this) can be found at \cite{website}. In the future, incremental improvements may be made to the look-up tables or interpolation code, but we consider this to be sufficient validation of our COM-Poisson code.

\bibliography{COMbib}

\end{document}